\documentclass[3p,times,twocolumn]{elsarticle}

\usepackage{ecrc}

\volume{00}

\firstpage{1}

\journalname{Nuclear Physics B Proceedings Supplement}

\runauth{}

\jid{nuphbp}

\jnltitlelogo{Nuclear Physics B Proceedings Supplement}

\usepackage{amssymb}
\usepackage[figuresright]{rotating}

\def\lr{\left( }
\def\rr{\right) }

\def\beq{\begin{equation}}
\def\eeq{\end{equation}}
\def\bea{\begin{eqnarray}}
\def\eea{\end{eqnarray}}

\begin{document}

\begin{frontmatter}

%% Title, authors and addresses

%% use the tnoteref command within \title for footnotes;
%% use the tnotetext command for the associated footnote;
%% use the fnref command within \author or \address for footnotes;
%% use the fntext command for the associated footnote;
%% use the corref command within \author for corresponding author footnotes;
%% use the cortext command for the associated footnote;
%% use the ead command for the email address,
%% and the form \ead[url] for the home page:
%%
%% \title{Title\tnoteref{label1}}
%% \tnotetext[label1]{}
%% \author{Name\corref{cor1}\fnref{label2}}
%% \ead{email address}
%% \ead[url]{home page}
%% \fntext[label2]{}
%% \cortext[cor1]{}
%% \address{Address\fnref{label3}}
%% \fntext[label3]{}

\dochead{}
%% Use \dochead if there is an article header, e.g. \dochead{Short communication}

\title{QCD analysis and effective temperature of direct photons \\ in lead-lead collisions at the LHC}

%% use optional labels to link authors explicitly to addresses:
%% \author[label1,label2]{<author name>}
%% \address[label1]{<address>}
%% \address[label2]{<address>}

\author{M. Klasen$^a$, F. K\"onig$^a$}

\address{$^a$Institut f\"ur Theoretische Physik, Westf\"alische
 Wilhelms-Universit\"at M\"unster, Wilhelm-Klemm-Stra\ss{}e 9,
 D-48149 M\"unster, Germany}

\author{C. Klein-B\"osing$^{b,c}$, J.P. Wessels$^b$}

\address{$^b$Institut f\"ur Kernphysik, Westf\"alische
 Wilhelms-Universit\"at M\"unster, Wilhelm-Klemm-Stra\ss{}e 9,
 D-48149 M\"unster, Germany}

\address{$^c$ExtreMe Matter Institute, GSI, Planckstra\ss{}e 1, D-64291 Darmstadt, Germany}

\begin{abstract}
%% Text of abstract
 We present a systematic theoretical analysis of the ALICE measurement
 of low-$p_T$ direct-photon production in central lead-lead collisions at the LHC
 with a centre-of-mass energy of $\sqrt{s_{NN}}=2.76$ TeV.  Using 
 next-to-leading order of perturbative QCD, we compute
 the relative contributions to prompt-photon production from
 different initial and final states and the theoretical uncertainties coming
 from independent variations of the renormalisation and factorisation scales,
 the nuclear parton densities and the fragmentation functions. Based on
 different fits to the unsubtracted and prompt-photon subtracted ALICE data,
 we consistently find an exponential, possibly thermal, photon spectrum from
 the quark-gluon plasma (or hot medium) with slope
  $T=304\pm 58$ MeV and $309\pm64$ MeV at $p_T\in[0.8;2.2]$ GeV and
 $p_T\in[1.5;3.5]$ GeV as well as a power-law ($p_T^{-4}$) behavior for $p_T>4$
 GeV as predicted by QCD hard scattering.
\end{abstract}

\begin{keyword}
%% keywords here, in the form: keyword \sep keyword

Quark-Gluon-Plasma \sep Photons \sep NLO QCD
%% MSC codes here, in the form: \MSC code \sep code
%% or \MSC[2008] code \sep code (2000 is the default)

\end{keyword}

\end{frontmatter}

%%
%% Start line numbering here if you want
%%
% \linenumbers

%% main text

\vspace*{-17cm}
\noindent MS-TP-14-29
\vspace*{15.5cm}

\section{Introduction}
\label{}

Direct photons, i.e.\ photons not originating from meson decay, represent an
important probe of the quark-gluon plasma (QGP) that is assumed to have existed
in the early Universe and that is currently under experimental investigation
in heavy-ion collisions at the CERN LHC. An extraction of the effective QGP
temperature from the transverse momentum ($p_T$) spectrum of the thermal photons
requires a precise knowledge and the subtraction of the prompt photons produced
in the primary hard collisions of quarks and gluons.

\begin{figure}
 \centering
 \includegraphics[width=\columnwidth]{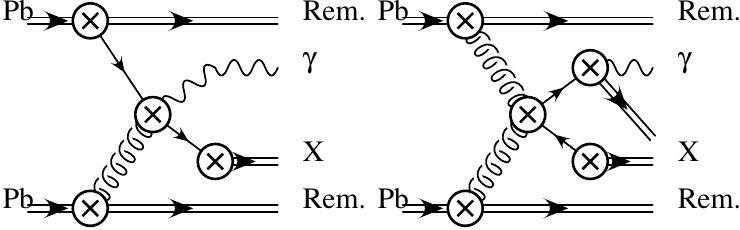}
 \caption{\label{fig:1}Feynman diagrams for prompt photon production through
 direct parton scattering (left) and fragmentataion processes (right).}
\end{figure}

These photons originate not only from the hard scattering processes $qg\to q\gamma$
(QCD Compton scattering) and $q\bar{q}\to g\gamma$ (quark-antiquark annihilation)
as shown in an exemplary way in Fig.\ \ref{fig:1} (left), but also from purely
%
%%%%%%%%%%%%%% Begin Figure 1 %%%%%%%%%%%%%%%%%%%%%%%%%%%%%%%%%%%%%%%%%%
%%%%%%%%%%%%%% End of Figure 1 %%%%%%%%%%%%%%%%%%%%%%%%%%%%%%%%%%%%%%%%%
%
partonic scattering processes followed by the fragmentation of a quark or gluon
into a photon as shown in Fig.\ \ref{fig:1} (right). At leading order (LO) of
perturbative QCD, these processes are formally of the same order ${\cal O}
(\alpha\alpha_s)$ due to the asymptotic ${\cal O}(\alpha/\alpha_s)$ dependence
of the photon fragmentation function (FF). At next-to-leading order (NLO),
they are related through the factorisation of logarithmic mass divergences
\cite{Klasen:2002xb}.

\section{Theoretical uncertainties}
\label{}

This procedure introduces a factorisation scale ($\mu_f$) dependence in the
real emission ($R$) squared scattering matrix elements through
\beq
 |{\cal M}^R|^2_{ab\to cde}=\ln\lr{\mu_f^2\over p_T^2}\rr|{\cal M}^B|^2_{ab\to cd}P_{d\to e}(x)
 +...,
\eeq
where $P_{d\to e}$ represents an Altarelli-Parisi splitting function. Together
with the renormalisation scale ($\mu$) dependence inherent in the strong coupling
constant $\alpha_s(\mu)$ through the QCD $\beta$ function and the virtual loop ($V$)
corrections,
\beq
 |{\cal M}^V|^2_{ab\to cd}=\ln\lr{\mu^2\over p_T^2}\rr|{\cal M}^B|^2_{ab\to cd}\beta_0 +...,
\eeq
it leads to a theoretical scale uncertainty that is typically estimated by independent
variations of $\mu$ and $\mu_f$ by factors of two around the central scale $p_T$.
\footnote{Additional uncertainties are induced by variations of the factorisation
{\em scheme} \cite{Klasen:1996yk}.}

\begin{figure}
 \centering
 \includegraphics[width=\columnwidth]{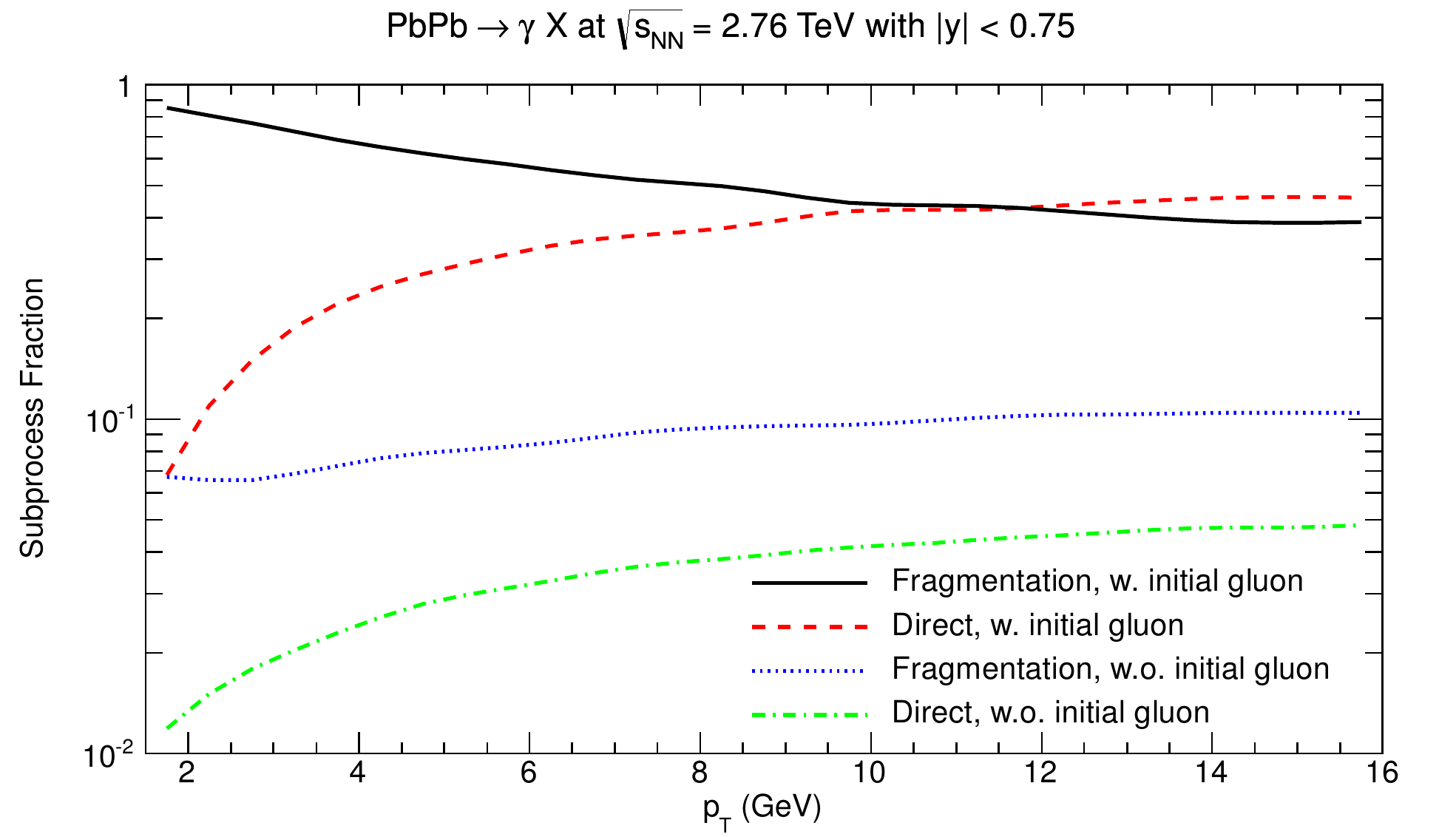}
 \caption{\label{fig:2}Relative contributions of direct and fragmentation
 processes initiated by quarks and gluons as a function of $p_T$.}
\end{figure}

Phenomenologically, fragmentation processes are usually important at low $p_T$, i.e.\
in the same region where the thermal photon signal is expected, whereas direct photons
dominate at high $p_T$. This is also the case in the ALICE experiment \cite{Wilde:2012wc}
performed with lead-lead collisions of $\sqrt{s}_{\rm NN}=2.76$ TeV at central rapidity
(cf.\ Fig.\ \ref{fig:2}).
%
%%%%%%%%%%%%%% Begin Figure 2 %%%%%%%%%%%%%%%%%%%%%%%%%%%%%%%%%%%%%%%%%%
%%%%%%%%%%%%%% End of Figure 2 %%%%%%%%%%%%%%%%%%%%%%%%%%%%%%%%%%%%%%%%%
%
Below $p_T=4$ GeV, direct photons contribute less than 25\%, fragmentation photons
more than 75\%. Since the FFs (in particular the one of the gluon) are not known
very precisely (i.e.\ with an uncertainty of up to an order of magnitude) \cite{Bourhis:1997yu}, 
similarly to the photon structure function \cite{Albino:2002ck}, they represent a second
important source of theoretical uncertainty. The reduction of this uncertainty, e.g.\
through dedicated measurements of low-$p_T$ inclusive photons in $pp$ collisions at RHIC,
is therefore quite important \cite{Klasen:2014xfa}. Alternatively, one may resort
to slightly off-shell photons which do not have important fragmentation contributions
\cite{Berger:1998ev}.

Fig.\ \ref{fig:2} demonstrates that gluon-initiated scatterings dominate
over those initiated by quarks in the whole $p_T$ range. The parton density
functions (PDFs) of lead nuclei, in particular the one of the gluon, are also not
precisely known, as they are affected by so-called cold nuclear effects such as shadowing
\cite{Eskola:2009uj}. This introduces a third source of theoretical uncertainty that
must be estimated, like the one for the FFs, by variations of different parametrisations
that describe the fitted experimental data equally well within one standard deviation.

\section{Results}
\label{}

\begin{figure}
 \centering
 \includegraphics[width=\columnwidth]{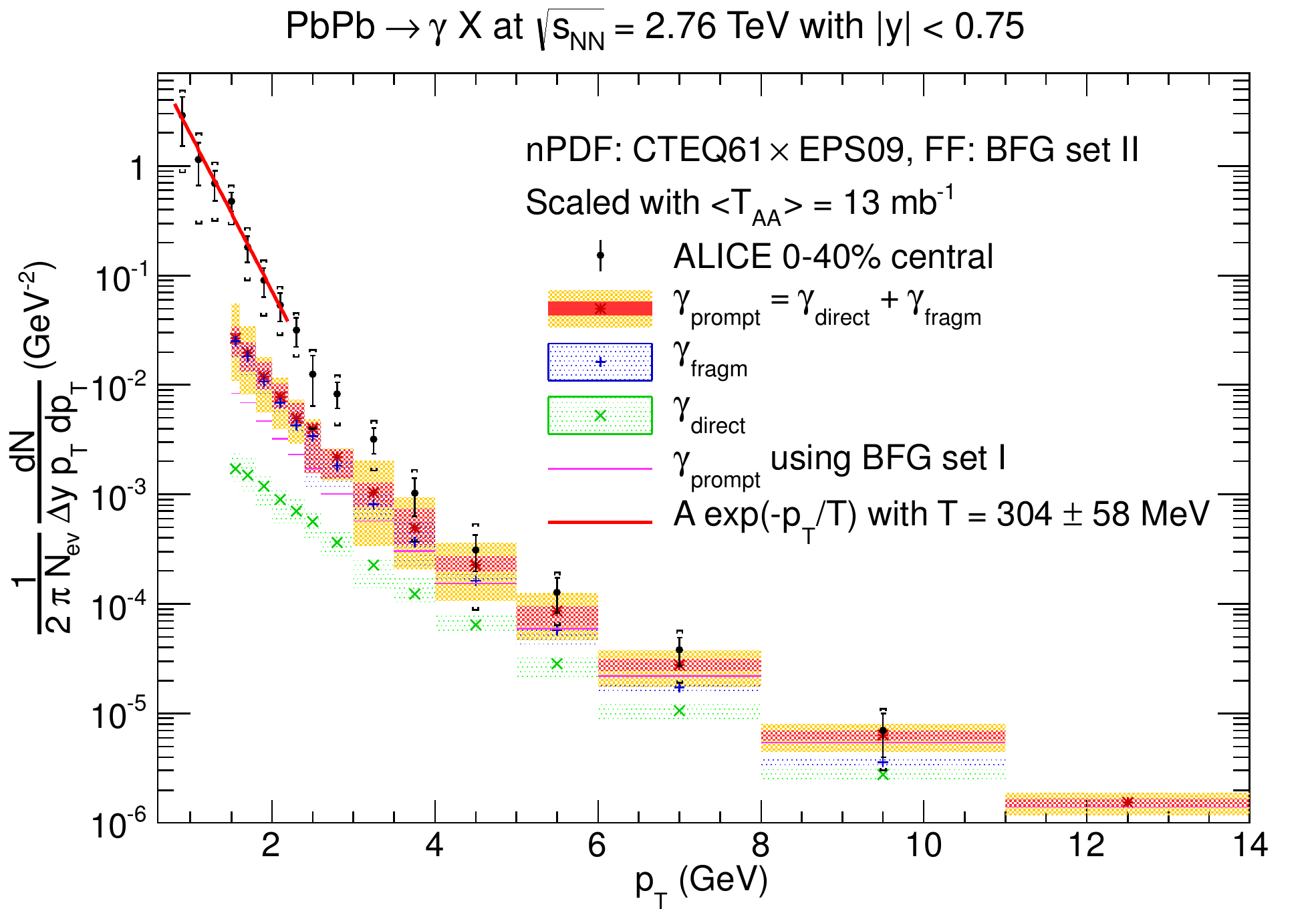}
 \caption{\label{fig:3}Transverse-momentum distribution of direct photons
 in central (0-40\%) lead-lead collisions at $\sqrt{s_{NN}}=2.76$ TeV
 (black data points \cite{Wilde:2012wc}) compared to our NLO QCD predictions
 for prompt photons (red) as well as their fragmentation (blue) and direct
 (green) contributions. Independent scale (yellow) and nPDF (red/blue/green)
 uncertainties as well as the FF variation (magenta) are shown separately, as 
 are the statistical (vertical bars) and systematic (horizontal brackets)
 experimental errors.}
\end{figure}

Using the NLO program JETPHOX \cite{Aurenche:2006vj}, we have computed all relevant
direct and fragmentation processes and compared them to the ALICE data \cite{Wilde:2012wc}
in Fig.\ \ref{fig:3} \cite{Klasen:2013mga}.
While we find very good agreement in the perturbatively dominated
%
%%%%%%%%%%%%%% Begin Figure 3 %%%%%%%%%%%%%%%%%%%%%%%%%%%%%%%%%%%%%%%%%%
%%%%%%%%%%%%%% End of Figure 3 %%%%%%%%%%%%%%%%%%%%%%%%%%%%%%%%%%%%%%%%%
%
large-$p_T$ bins, the ALICE data show a clear excess below 4 GeV. Even though
the FF uncertainty is indeed very large, it can still not account for the data.
The scale (yellow) and PDF (red) uncertainties are also quite substantial.
With an exponential fit
\beq
 A\exp(-p_T/T)
\eeq
to the region 0.8 GeV $<p_T<$ 2.2 GeV, we extract an inverse slope parameter $T$
of 304$\pm58$ MeV, that is very similar to the value 304$\pm$51 MeV obtained
by the ALICE collaboration \cite{Wilde:2012wc}. However, the error is now
larger as it includes also the various theoretical sources of uncertainty
described above.

\begin{figure}
 \centering
 \includegraphics[width=\columnwidth]{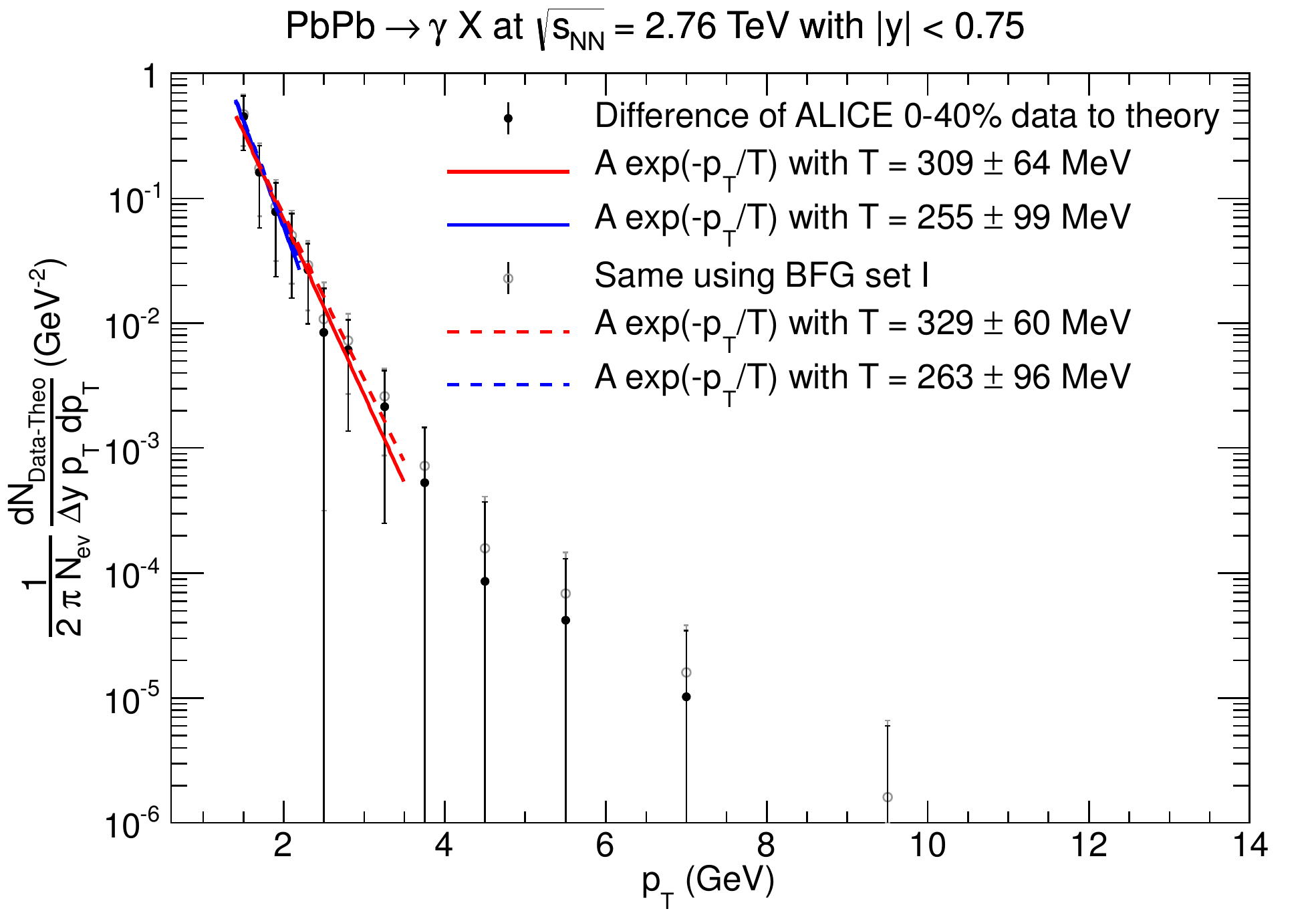}
 \caption{\label{fig:4}Difference of experimental direct-photon and
 theoretical prompt-photon yields using BFG II (full circles)
 and BFG I (open circles) FFs with all other errors added in quadrature.
 Also shown are exponential fits to the points up to $p_T<2.2$ GeV (blue)
 and 3.5 GeV (red).}
\end{figure}

Since the $p_T$ region selected in this fit is somewhat arbitrary, we perform in
Fig.\ \ref{fig:4}
%
%%%%%%%%%%%%%% Begin Figure 4 %%%%%%%%%%%%%%%%%%%%%%%%%%%%%%%%%%%%%%%%%%
%%%%%%%%%%%%%% End of Figure 4 %%%%%%%%%%%%%%%%%%%%%%%%%%%%%%%%%%%%%%%%%
%
two additional fits to the ranges 1.5 GeV $<p_T<$ 2.2 GeV and 1.5 GeV $<p_T<$ 3.5
GeV. This can, however, only be done after properly subtracting the perturbative
contribution, which starts to be substantial at intermediate $p_T$. The latter
fit is of course more reliable, as it includes more data points, and leads to
values of $T=309\pm64$ MeV and 329$\pm$60 MeV using BFG I and II FFs, respectively.
These values are again similar to those above.

\begin{figure}
 \centering
 \includegraphics[width=\columnwidth]{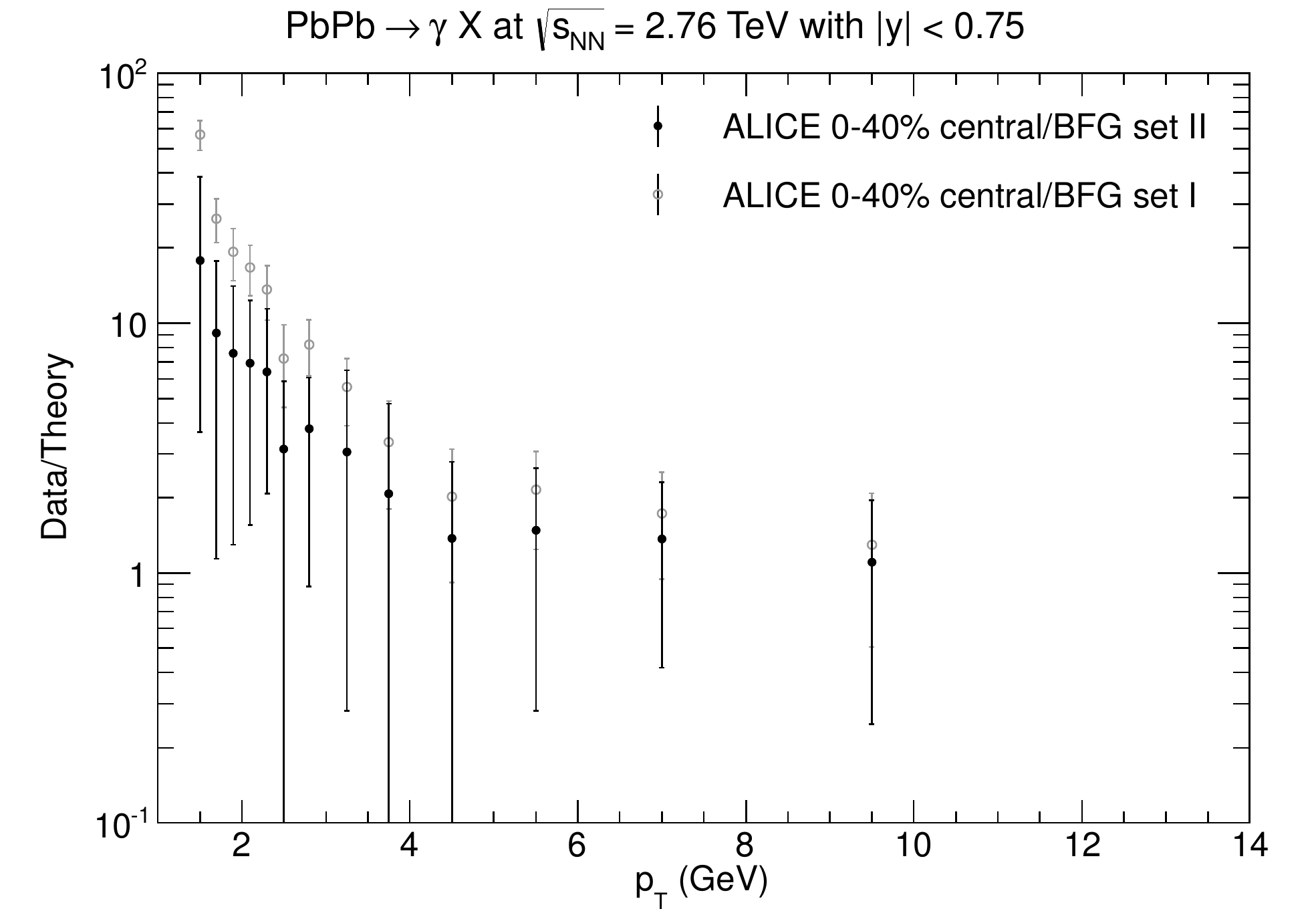}
 \caption{\label{fig:5}Ratio of experimental direct-photon and
 theoretical prompt-photon yields using BFG II (full circles)
 and BFG I (open circles) FFs with all other errors added in quadrature.}
\end{figure}

This result may at first sight seem surprising and coincidental.
One must, however, take into account the relative size of the
prompt-photon contribution, which falls off like $p_T^{-4}$
as predicted by our calculations, to
the measured direct-photon rate in the different $p_T$ regions,
shown in Fig.\ \ref{fig:5}. Indeed, we find that prompt photons
%
%%%%%%%%%%%%%% Begin Figure 5 %%%%%%%%%%%%%%%%%%%%%%%%%%%%%%%%%%%%%%%%%%
%%%%%%%%%%%%%% End of Figure 5 %%%%%%%%%%%%%%%%%%%%%%%%%%%%%%%%%%%%%%%%%
%
contribute less than 20\% for $p_T<2.4$ GeV (less than 10\% for
BFG I FFs), i.e.\ their subtraction and the theoretical error
do not strongly modify the exponential fall-off of the
experimental data in this region. Above 4 GeV,
we find instead an almost constant ratio of prompt over direct
photons that is consistent with one within the uncertainties,
i.e.\ these observed photons are produced in hard scatterings.
In the intermediate region, it was already clear from Fig.\
\ref{fig:3} that the prompt-photon contribution is substantial,
larger than 20\% (10\% for BFG I FFs), and must indeed first be
subtracted from the experimental data. Only then can the excess
be described again by an exponential fit, which we found to be
in good agreement with the fit to the unsubtracted data at low
$p_T$.

\section{Summary}
\label{}

In conclusion, we have performed in this work a first systematic theoretical
analysis of the ALICE measurement of direct-photon production in central
lead-lead collisions with a centre-of-mass energy of
$\sqrt{s_{NN}}=2.76$ TeV at low values of $p_T$ of
0.8 to 14 GeV. Based on a next-to-leading order QCD calculation,
we found that prompt photon production in this region was
largely induced by initial gluons and dominated by fragmentation
contributions. This resulted in large theoretical uncertainties
from independent variations of the renormalisation and factorisation
scales, nuclear parton densities and fragmentation functions.

Nevertheless, we were able to confirm that the experimental data are well fitted
in the region $p_T\in[0.8;2.2]$ GeV by an exponential form $A\,\exp(-p_T/T)$
with an effective temperature of $T=304$ $\pm$ 58 MeV, with an only slightly
larger uncertainty than the 51 MeV quoted by the ALICE collaboration.
The reason is that in this region  prompt photons contribute
less than 10 to 20\% to the total direct photon rate,
so that their subtraction and theoretical error
do not influence the fit result very much. We also verified that already for
values of $p_T>4$ GeV the experimental data fall off with approximately
$p_T^{-4}$ as predicted by perturbative QCD.
In the intermediate $p_T$-region from 1.5 to 3.5 GeV, the prompt-photon
contribution had to be subtracted from the experimental data before
a sensible exponential fit could be performed. We were able to verify
an exponential fall-off with a very similar effective temperature
of $309\pm64$ MeV.
The inverse slope parameter of this
measurement is significantly higher than the one obtained previously
by the PHENIX collaboration in 0-20\% central gold-gold collisions with
$\sqrt{s_{NN}}=200$ GeV at RHIC, which resulted in $T_{\rm RHIC}=221\pm19\pm19$ MeV
\cite{Adare:2008ab}. The latter was higher than
the transition temperature to the QGP of about $T_{\rm crit}=170$ MeV, but
1.5 to 3 times smaller
than the initial temperature $T_0$ of the dense matter due to the
space-time evolution following its initial formation;
in hydrodynamical models, which describe the data at $\sqrt{s_{NN}}$ =
200 GeV, $T_0$ ranges from 600 to 300 MeV depending on the
formation time, assumed to lie between $\tau_0 =  0.15$ and
0.6 fm$/c$ \cite{d'Enterria:2005vz}.

\section*{Acknowledgments}

The work of C.\ Klein-B\"osing was supported by the Helmholtz Alliance
Program of the Helmholtz Association, contract HA216/EMMI ``Extremes of
Density and Temperature: Cosmic Matter in the Laboratory''.
We thank the ALICE collaboration for making their preliminary
data available to us and J.P.\ Guillet, J.\ Owens, W.\ Vogelsang
and M.\ Wilde for useful discussions.

%% The Appendices part is started with the command \appendix;
%% appendix sections are then done as normal sections
%% \appendix

%% \section{}
%% \label{}

%% References
%%
%% Following citation commands can be used in the body text:
%% Usage of \cite is as follows:
%%   \cite{key}         ==>>  [#]
%%   \cite[chap. 2]{key} ==>> [#, chap. 2]
%%

%% References with BibTeX database:
%\nocite{*}
%\bibliographystyle{elsarticle-num}
%\bibliography{martin}

%% Authors are advised to use a BibTeX database file for their reference list.
%% The provided style file elsarticle-num.bst formats references in the required Procedia style

%% For references without a BibTeX database:

\end{document}